\documentclass[notitlepage, twocolumn,floatfix,superscriptaddress,prd]{revtex4-1}
\usepackage{graphicx, epsfig, bm, amsmath}
\usepackage{amssymb}
\usepackage{amsmath}
\usepackage{color}
\usepackage{wasysym}
\usepackage{savesym}
\usepackage{hyperref}

\def\barray{\begin{array}}
\def\earray{\end{array}}
\def\be{\begin{equation}}
\def\ee{\end{equation}}
\def\ben{\begin{equation} \nonumber}
\def\een{\end{equation}}
\def\ban{\begin{eqnarray*}}
\def\ean{\end{eqnarray*}}
\def\ba{\begin{eqnarray}}
\def\ea{\end{eqnarray}}

\def\curv{\mathcal{R}}

\def\({\left(}
\def\){\right)}
\def\[{\left[}
\def\]{\right]}
\def\half{{1\over2}}
\def\tr{{\rm Tr}}
\def\nn{\nonumber}
\def\hdaxion{\hat{X}}

\def\mpsi{m_{\psi}}

\def\tr{{\rm Tr}}
\def\hg{{\hat\gamma}}
\def\hy{{\hat t}}
\def\unhg{\gamma}
\def\unhy{t}

\def\MM{{\cal M}}
\def\GG{{\cal G}}

\def\AA{{\cal A}}

\def\One{{\hbox{ 1\kern-.8mm l}}}

\def\ada{{\hat X}}
\def\axsub{{\hat x}}

\def\phitil{{\boldsymbol\varphi}}
\def\phitilb{{\bar{\boldsymbol\varphi}}}
\def\ztil{{\boldsymbol z}}
\def\ztilb{{\bar{\boldsymbol z}}}

\newcommand {\bflam}{{\boldsymbol\Lambda}}

\def\da{\Psi}

\def\axion{\mathcal{X}}
\def\daxion{\delta\!\mathcal{X}}

\definecolor{darkgreen}{cmyk}{0.85,0.2,1.00,0.2}

\newcommand{\pref}[1]{$(\ref{#1})$}

\def\be{\begin{equation}}
\def\ee{\end{equation}}
\def\bea{\begin{eqnarray}}
\def\eea{\end{eqnarray}}

\begin{document}

\title{Gauge Fields and Inflation: Chiral Gravitational Waves, \\ Fluctuations and the Lyth Bound} %
\author{Peter Adshead}
\affiliation{Kavli Institute for Cosmological Physics, University of Chicago, Chicago, Illinois 60637, U.S.A}
\affiliation{Enrico Fermi Institute, University of Chicago, Chicago, Illinois 60637, U.S.A}
\author{Emil Martinec}
\affiliation{Enrico Fermi Institute, University of Chicago, Chicago, Illinois 60637, U.S.A}
\affiliation{Department of Physics, University of Chicago, Chicago, Illinois 60637, U.S.A.}
\author{Mark Wyman}
\affiliation{Kavli Institute for Cosmological Physics, University of Chicago, Chicago, Illinois 60637, U.S.A}  
\affiliation{Enrico Fermi Institute, University of Chicago, Chicago, Illinois 60637, U.S.A}
\affiliation{Department of Astronomy \& Astrophysics, University of Chicago, Chicago, Illinois 60637, U.S.A.}

\begin{abstract}
Models of inflation involving non-Abelian gauge field backgrounds can produce gravitational waves at an observable level with a preferred handedness. 
This asymmetry comes about because the non-Abelian
background generates parity-violation in the action for perturbations. 
In the specific model we study, Chromo-Natural Inflation, these gravitational waves
can be produced at observable levels even when no field makes a super-Planckian field excursion, thus evading a common formulation of the Lyth bound.
Unfortunately, when considered in concert with the scalar fluctuations, this chiral enhancement of  the gravitational waves makes the model observationally inviable.
\end{abstract}

\maketitle

Scalar fields generically drive the inflationary epoch. However, some of the qualitative problems of inflationary
model building can be ameliorated by incorporating gauge fields into the dynamics~\cite{Adshead:2012kp,Adshead:2012qe,Maleknejad:2011sq, Anber:2009ua,Martinec:2012bv}.   In particular, non-Abelian gauge
fields can appear in a homogeneous and isotropic configuration, leading to slow-roll inflation even on steep potentials~\cite{Adshead:2012kp} via competition between the potential forces and magnetic drift forces induced by a Chern-Simons interaction~\cite{Martinec:2012bv}. In this work, we study the perturbations in such models, and
show that the gauge field background results in a parity-violating action for fluctuations.  During inflation, this parity violation imprints on the spectrum of gravitational perturbations; one helicity is enhanced relative to the other, leading to a classically generated chiral spectrum of tensor fluctuations. We note that chiral gravitational wave production from quantum effects was found in a related context in Refs. ~\cite{Sorbo:2011rz,Anber:2012du}. The amplitudes of the gravitational waves (in either polarization)
are not directly correlated to the (sub-Planckian) excursion of the inflaton field during inflation, even though this is a single-clock inflationary setting, because the motion of the inflaton along its potential is mainly resisted by an analog of magnetic drift, not Hubble friction. Thus, this mechanism evades the simplest version of the so-called Lyth bound \cite{Lyth:1996im}. Unfortunately, the efficient production of gravitational waves leads to it being impossible for the model as currently written to agree with observations.
A more comprehensive presentation of our analysis will appear in a forthcoming paper \cite{Adshead:2013nka}. In this work, we adopt natural units: $M_{\rm Pl} = c = \hbar =1$. 

\section{Non-Abelian Gauge background:}

The Einstein-Yang-Mills equations are solved on FRW for a non-Abelian gauge field with this configuration:
\begin{align}\label{gaugevev}
A_{0} =  0,\; A_{i} =  \phi\delta^{a}_{i} J^{a} \equiv a\psi \delta^{a}_{i} J^{a} \\
 F_{0i} =  \partial_{\tau}\phi \delta^{a}{}_{i}J^a, \;
F_{ij}  =  g \phi^2 f^{a}_{ij} J^a.
\end{align}
where $J_a$ is a generator of SU(2) %
\footnote{Other embeddings of $SU(2)\subset SU(N)$ are possible;  we take $N=2$ for simplicity. 
We also take $\tr[J_a J_b] \equiv \half \delta_{ab}$.}.
This gauge field background violates parity by relating space and gauge indices with a particular orientation. Perturbations around this background thus exhibit parity-violating interactions.  We will study the model of Chromo-Natural Inflation, which is composed of an axion,  $\axion$, interacting with the gauge fields via
\be
{\cal L}
 =  \frac{1}{2}R -\frac{1}{2}(\partial\axion)^2 -V(\axion)-\frac{1}{2}\tr[ F^2] \!-\frac{\lambda}{4 f}\axion \tr[ F\wedge F]  .
\ee 
In \cite{Adshead:2012kp} we showed that this model inflates, with slow-roll provided via a magnetic-drift type force
mediated by the Chern-Simons interaction.  In the large drift force limit ($\lambda \gg 1$), the slow-roll equations for this model are
\begin{align}
\dot \psi & = - H \psi + \frac{f }{3 g \lambda} \frac{V_{,\axion}}{\psi^{2}}  
~,\quad \label{slowrollpsi}
\frac{\lambda}{f} \dot \axion 
 = 2 g \psi +  \frac{2H^2}{g \psi}  ~,
\end{align}
where $g$ is the gauge coupling, $f$ is the axion decay constant and  here and throughout an overdot denotes a derivative with respect to cosmic time. We also take $V=\mu^4 (1+\cos (\axion/f))$. Slow roll solutions of this system occur at large values of the parameter $\lambda \sim \mathcal{O}(10^2 - 10^4)$, and yield approximately static gauge field solutions where $\psi \sim 10^{-2}$ and $\dot\psi \approx \mathcal{O}(\psi^2)$. For the remainder of this paper, we will  take $\dot\psi = 0$, consistent with the slow-roll approximations made throughout.

We work with conformal time, $d \tau \equiv dt/a$ which we define to be a negative quantity during inflation with $\tau = 0$ corresponding to the end of inflation. It will also prove useful to introduce the dimensionless 
time $x = -k \tau$, where $k$ is the wavenumber of the mode under consideration. For comparison with data, we will take $k$ to have cosmological units, $h$/Mpc, and
this will fix units for $\tau$. Finally, we will
use throughout two derived background variables:
\be
m_\psi \equiv \frac{g \psi}{H}, \qquad \quad \bflam \equiv \lambda \frac{\psi}{f},
\ee
where $m_\psi$ is dimensionless and characterizes the mass of the gauge field fluctuations in units of the Hubble scale. Note also that $\psi \sim {\cal O}(f)$, so $\bflam \sim {\cal O}(\lambda)$, which we will take to be large.

\looseness-1
\section{Perturbations overview}
Before delving into the details, let us sketch the novel features of the perturbations in this scenario. Because we have a non-Abelian gauge
field on a background, we will have additional propagating scalar, vector, and tensor modes.  The perturbative dynamics exhibits standard behavior at early times $-k\tau \to \infty$, where as usual one recovers free plane waves; also, for late times $-k \tau \to 0$, the fluctuations freeze out, leaving only a single growing scalar and a single growing tensor mode.  There are, however, two novel behaviors at intermediate times:

\begin{enumerate}
\item For $\bflam\mpsi \gtrsim |k \tau| \gtrsim \mpsi$, we find a complex interacting system for the scalars.  The mode that governs the amplitude of the axion at superhorizon scales decays by a factor of order $\bflam$.  This reduced amplitude is compensated by an enhancement of order $\bflam$ in the relation between the amplitudes of the inflaton and scalar curvature perturbations.
\item For a period $k\delta\tau\approx\mpsi$ around $|k\tau| \approx 1$, one polarization of the gauge field tensor modes becomes unstable and grows exponentially.  This mode sources the corresponding gravitational tensor polarization, enhancing it relative to the other polarization.  
\end{enumerate}
For the present work, we will ignore the vector perturbations which do not play an important role; unlike the tensors, their masses are never strongly tachyonic, and the (mild) amplification that they feel near horizon crossing
is observationally irrelevant since all vector modes damp at late times \cite{Adshead:2013nka}.

\section{Some Details}
We will work with the metric in ADM form, 
\begin{align}\label{eqn:adm}
ds^2 = -N^2 d\tau^2 +  \tilde{h}_{ij}(dx^i+N^i d\tau)(dx^j+N^j d\tau),
\end{align}
where $N$ is the lapse,  $N^i$ is the shift vector and $\tilde{h}_{ij}$ is the metric on spatial hypersurfaces. In our conventions, on the FRW background the lapse and shift have solutions $N = a$ and $N^i = 0$. We choose spatially 
flat gauge, in which
 (with $\gamma_{ij}$ transverse-traceless)\footnote{In our conventions, all gauge indices and repeated lower indices are summed with the Kronecker delta. Paired upper and lower indices  are summed using the spacetime metric $g_{\mu\nu}$}
\begin{align}\label{eqn:spatialmet}
\tilde{h}_{ij} = a^2\[e^{\gamma}\]_{ij} = a^2\left[\delta_{ij} + \gamma_{ij} + \frac{1}{2}\gamma_{im}\gamma_{mj}+\ldots\right].
\end{align}
We write the fluctuations of the gauge fields as $\delta\! A_\mu \equiv \Psi_\mu$ and decompose their spatial parts as:
\be
\label{SVTdecomp}
 \da_i= (t_i^a+\epsilon_{ik}^a\chi_k + \delta_i^a\delta\!\phi)J_a \ ,
\ee
where $t$ is symmetric and traceless, $\chi_k$ is a general vector, and the axion perturbs to $\axion + \daxion$. 
In what follows, we work with explicit components of the fields by choosing the wavenumber along the $x^3$ direction. 
The gauge modes~\pref{SVTdecomp} then have an SVT decomposition in which
 \be
 t^\pm = \frac12(t_{11}-t_{22})\pm i t_{12} \ee are the polarizations of a transverse traceless tensor, 
 \be
  t_{3,(1\pm i2)}, \quad  \rm{and} \quad \chi_{1\pm i2},
 \ee
are the four polarizations of two transverse vectors, and 
\be
z\equiv\frac16(2t_{33}-t_{11}-t_{22}), \quad \rm{and} \quad \chi_3,
\ee
 are scalars along with $\delta\!\phi$.

\subsection{Tensors}
We first consider the behavior of spin-2 modes of the metric and gauge fields. The spin-2 modes of the metric, $\gamma_{ij}$, the gravitational waves,  are invariant under coordinate transformations at linear order and are not subject to the linear order Einstein constraints. On the other hand, at linear order the spin-2 modes of the gauge field are invariant under SU(2) gauge transformations and further are not subject to the linear Gauss's law constraint.

The quadratic action for these tensor modes is:
\begin{align}\nn
\mathcal{S}_T  &=  \frac{1}{2}\int \frac{d^3 k}{(2\pi)^3} d\tau \Bigl[
2\partial_{\tau}\unhy_k^{\pm}\partial_{\tau}\bar{\unhy}_k^{\pm}
- 2\Bigl(k^2 +  g\phi\frac{\lambda}{2f}\partial_{\tau}\axion  \Bigr)\unhy_k^{\pm}\bar{\unhy}_k^{\pm}\\
\nn  \pm & 2k\Bigl(\frac{\lambda}{f}\partial_{\tau}\axion 
+ 2g\phi\Bigr) \unhy_k^{\pm}\bar{\unhy}_k^{\pm}
-  2\dot{\phi}( \partial_{\tau} \unhy^{\pm}_k\bar{\unhg}_k^{\pm}
+ \partial_{\tau} \bar{\unhy}_k^{\pm}{\unhg}_k^{\pm})\\ 
\mp & 2k g\frac{\phi^2}{a} (\unhy_k^{\mp}\bar{\unhg}_k^{\pm}+\bar{\unhy}_k^{\pm}{\unhg}_k^{\pm}) 
+ 2g^2\frac{\phi^3}{a} (\unhy_k^{\pm}\bar{\unhg}_k^{\pm}+\bar{\unhy}_k^{\pm}{\unhg}_k^{\pm})
\label{eqn:tensoraction} \\
+ & \frac{1}{2}\partial_{\tau}\unhg_k^{\pm}\partial_{\tau}\bar{\unhg}_k^{\pm}
- \frac{1}{2}\Bigl(k^2-\frac{1}{a}\frac{\partial^2 a}{\partial\tau^2}
- 2\dot{\phi}^2 + 2 g^{2}\frac{\phi^4}{a^2}\Bigr)\unhg_k^{\pm}\bar{\unhg}_k^{\pm} \Bigr], \nn
\end{align}
where   $ \gamma^\pm/a$ and $\unhy^{\pm}$  are the amplitudes of left- and right-handed gravitational waves and spin-2 fluctuations of the gauge field, respectively. 
The canonically normalized fields are thus
$
{\hg}^{\pm} ={\unhg^{\pm}}/{\sqrt{2}}
$ 
and
$
{\hy}^{\pm} =  \sqrt{2} \unhy^{\pm}.
$
The crucial helicity dependence comes in the first term on the second line, which generates a tachyonic mass over a range of scales near horizon crossing for $t^{+}$ but not for $t^{-}$, and leads to a substantial growth of the former for a time near $x\sim \mpsi$ (see Fig.\ \ref{fig:analsol} below). During this growth, the coupling between the gravitational and gauge modes (which also has a mild chirality dependence) feeds the disparate amplitudes of the gauge tensor modes into the gravitational sector. It is worth emphasizing that this chiral splitting of tensor mode amplitudes is \emph{not} solely due to the Chern-Simons interaction.
The helicities are still split even in the absence of the Chern-Simons interaction ($\lambda \to 0$), because of the preferred orientation implicit in the gauge field background.  In this specific example, the absence of the Chern-Simons interaction would cause the parity-violating background to quickly decay away. However, any non-Abelian inflationary model relying on a non-trivial gauge texture of the form~\pref{gaugevev} will produce a chiral spectrum of tensor modes %
\footnote{
The absence of parity violation in the effective action of v1 of ref.~\cite{Dimastrogiovanni:2012st} indicates that some important effects were missed in that analysis.  The premise of this work is that the gauge modes are heavy and may be integrated out for $\mpsi\gg 1$, while in fact there is in this limit a substantial period of time when certain gauge modes are not only light but unstable.
}.

\begin{figure}[t]
\includegraphics[width = 3.5 in]{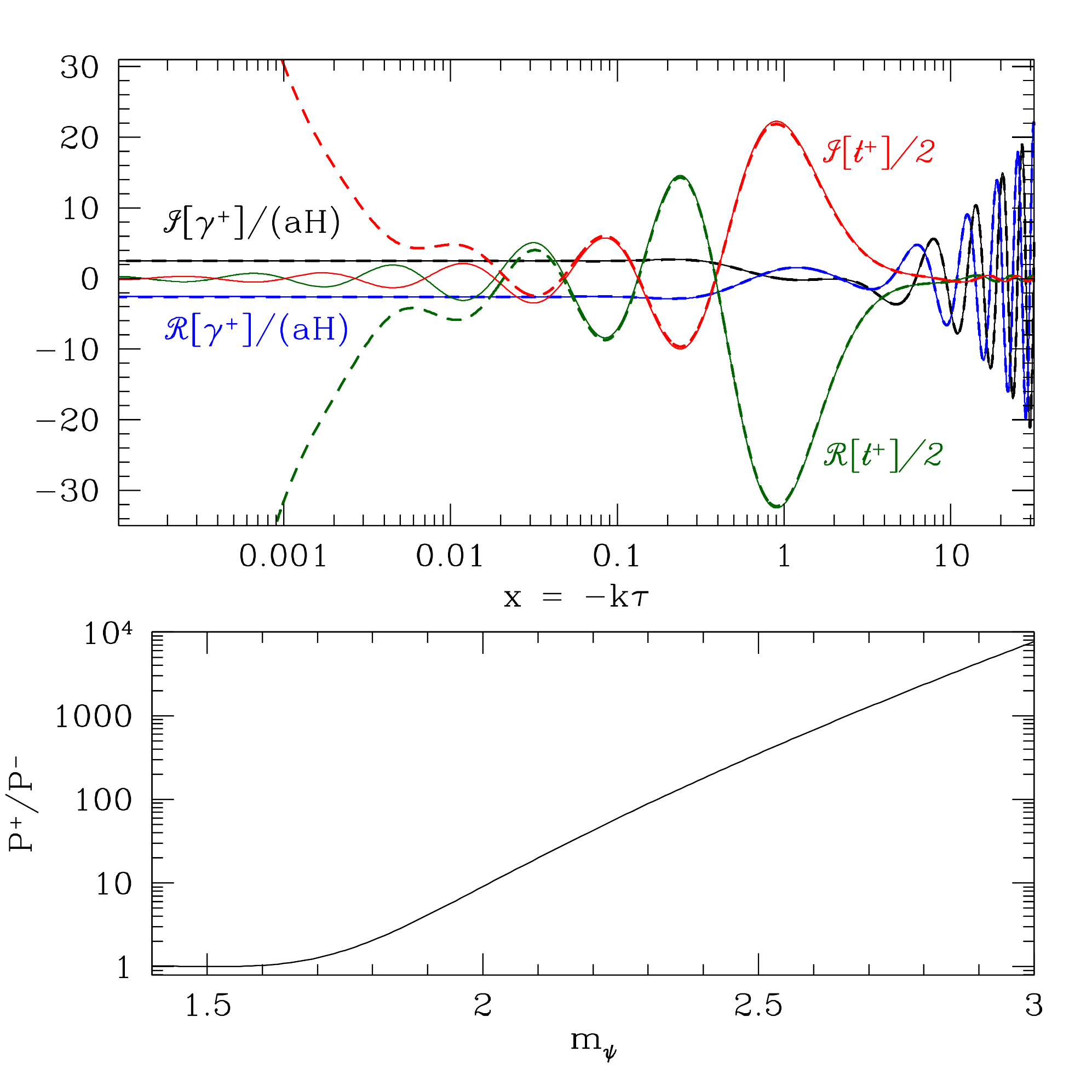}
\caption{
In the upper panel, we plot in blue and black curves the real and imaginary parts, respectively, of the physical left-handed gravitational wave perturbation $\gamma^+/a$ in units of the Hubble rate. In green and red curves we plot the real and imaginary parts of the physical left-handed spin-2 gauge field fluctuations $t^+$.  In dashed lines show the numerical integration of the equations resulting from the action, Eqn.~(\ref{eqn:tensoraction}). Solid lines show the result of the approximation of Eqns.\ (\ref{eqn:metricfull}) and (\ref{eqn:gaugeapprox}). The values of the parameters are chosen so that $m_{\psi} \approx 2.1$ and $\psi \approx 0.048$. The lower panel shows the ratio of the power in the two gravitational wave helicities as a function of $m_\psi$ holding $\psi$ fixed.}\label{fig:analsol}
\end{figure}

Working in the quasi de Sitter limit where $H\approx {\rm const}$, $a \approx -1/H\tau$, and $\dot{\phi} \approx H\phi$, and introducing the variable $x = - k \tau$, the equation of motion for the gravitational wave can be written 
\footnote{To modify this analysis to encompass the gauge-flation model~\cite{Maleknejad:2011sq}, simply replace $(\lambda/f) \partial_\tau \axion/a$ with $\kappa H \epsilon_H g \psi^3/2$.}
\begin{align}\nn
\label{eqn:metricfull}
\hg_k^{\pm}{}''+\Big(1-\frac{2}{x^2}-&2\frac{\psi^2}{x^2} 
+ 2 \frac{\mpsi^2\psi^2}{x^2}\Big)\hg_{k}^{\pm} \\
= & 2 \frac{ \psi}{x} \hy^{\pm}_k{}'
+ 2\psi \frac{\mpsi}{x^2}(\mpsi \mp x)\hy^{\pm}_k 
\end{align}
where here and throughout primes denote derivatives w.r.t.\ $x$. The gauge field satisfies the approximate equation
\begin{align}\label{eqn:gaugeapprox}
\hy^{\pm}_{k}{}''+\(1 + \frac{m}{ x^2}  \mp \frac{m_t}{x} \)\hy^{\pm}_k = & 0.
\end{align}
where we have defined%
\bea\label{eqn:mtdef}\nn
m & = &  \mpsi
\frac{\lambda}{f}\frac{\dot{\axion}}{H}
= 2(1+\mpsi^2) = \frac{1}{4}-\beta^2
\\
m_t &=&  \frac{\lambda}{f}\frac{\dot{\axion}}{H}
+2 \mpsi
=2(1+2\mpsi^2)/\mpsi  = -2i \alpha ~.
\eea
In Fig.\  \ref{fig:analsol}, we plot the solutions of a numerical evaluation of the coupled system of equations that results from the action in Eqn.\ (\ref{eqn:tensoraction}) in dashed lines. To demonstrate the accuracy of the approximation made to arrive at Eqn.\ (\ref{eqn:gaugeapprox}), we show the result of numerical evaluation of the complete system in solid lines. Plotted is the evolution of the amplitudes of the physical field fluctuations of the left-handed gravitational waves and gauge field fluctuations where both modes start off as plane waves at early times, $x \to \infty$.  For a period of time near $x\sim \mpsi$ the gauge field mass becomes temporarily negative leading to a period of exponential growth. After horizon crossing, compensating positive mass terms grow faster than the tachyonic terms, shutting off the instability and  leading to oscillatory decay as $x \to 0$. We note that in the region where the approximation at (\ref{eqn:gaugeapprox}) is breaking down (see the dashed lines in the upper panel of Fig. \ref{fig:analsol}), the metric fluctuations have already frozen out. Hence, the spurious growth of the gauge tensor is irrelevant to observable gravitational waves, which justifies our use of it %
\footnote{Note, however, that the linearized equations for graviton production break down as $\mpsi$ becomes large, as $t^+$ becomes exponentially large, and nonlinear effects such as $t^+t^+\rightarrow\daxion$ become important.}.

In the approximation, Eqn.\ (\ref{eqn:gaugeapprox}), the equation of motion for $\hy^+$ has an exact solution:
\begin{align}
\hy_k(x) = &  A_k M_{\alpha,\beta}(2 i x)+B_k W_{\alpha,\beta}(2 i x),
\end{align}
where $M_{\alpha,\beta}(2 i x)$ and $W_{\alpha,\beta}(2 i x)$ are the Whittaker M and W functions and $\alpha$, and $\beta$ were defined in Eqn.\ (\ref{eqn:mtdef}). We set  $A_k$ and $B_k$, by demanding that the solutions approach canonically normalized positive frequency free plane waves as $x \to \infty$:
$\hg^{\pm}, \hy^{\pm} \to e^{i x}/\sqrt{2k}.
$
We find
\begin{align}
A_k =\frac{(2 i)^\alpha  {\cal C} }{\Gamma(2\beta+1)}, \; B_k = \frac{2^\alpha i^{\beta+1} (-i)^{\alpha+\beta}  {\cal C}}{\Gamma(\alpha+\beta+\half)},
\end{align}
where ${\cal C} \equiv \Gamma(-\alpha+\beta+\half)/\sqrt{2k}$.
We can now solve for the  resulting positive helicity gravitational wave mode by inverting Eqn.\ (\ref{eqn:metricfull}).
The resulting integrals can be evaluated in closed form in terms of Meijer G-functions. The late time result
of this approximation scheme is
\begin{align}
\label{enhancement}
\hg^{+}(x)= &\hg_0(x)+\frac{2\psi }{x}B_k \Bigl(I_1+m_\psi \( I_2 - m_\psi I_3 \)\Bigr)
\end{align}
where $\hg_0(x)$ is the homogeneous  solution and
\begin{align}\nn
I_1 = &-i \frac{\left(m^2-2 i m m_t+2 m-2 m_t^2\right) \sec \left( \pi \beta \right) \sin \left( \pi\alpha\right) \Gamma \left(\alpha\right)}{2 m (m+2)} \\\nn &-i\frac{\pi ^2 \left(m^2+2 i m m_t+2 m-2 m_t^2\right) \sec \left( \pi \beta \right) \text{csc}\left(\pi\alpha\right)}{2 m (m+2) \Gamma \left(\alpha+1\right) \Gamma \left(\frac{1}{2}-\alpha-\beta\right) \Gamma \left(\frac{1}{2}-\alpha+\beta\right)}, \\\nn
I_2 = & \frac{\pi  \sec \left(\pi \beta\right) \Gamma \left(-\alpha\right)}{2 \Gamma \left(\frac{1}{2}-\alpha-\beta\right)
   \Gamma \left(\frac{1}{2}-\alpha+\beta\right)} +\frac{\pi ( m -i   m_t )\sec \left( \pi  \beta\right)}{2 m \Gamma \left(1-\alpha\right)} \nn \\ \nn& 
   -\frac{\pi  \sec \left(\pi \beta\right) \Gamma \left(1-\alpha\right)}{m \Gamma \left(\frac{1}{2}-\alpha-\beta\right) \Gamma \left(\frac{1}{2}-\alpha+\beta\right)}, \\ 
\nn I_3 = &i\frac{\pi ^2 (m+i m_t) \text{sec} \left(\pi 
   \beta\right) \text{csc}\left(\pi\alpha \right)}{m (m+2) \Gamma \left(\alpha \right) \Gamma \left(\frac{1}{2}-\alpha-\beta\right) \Gamma \left(\frac{1}{2}-\alpha+\beta\right)}\\  & +  \frac{\pi  (m_t+i m) \text{sec} \left( \pi\beta\right)}{m (m+2) \Gamma \left(-\alpha\right)} ~.
\end{align}

The right handed modes are unaffected by the gauge field, and have power $\Delta^{2}_{\gamma^-}(k) =  H^2/\pi^2$ while the left handed  gravitational wave power is given by
\begin{align}\label{eqn:GWpower}
\Delta^{2}_{\gamma^+}(k) = & \frac{H^2}{\pi^2}+4\frac{H^2}{\pi^2}\psi^2  k |B_k|^2 |I_1+m_\psi \( I_2 - m_\psi I_3 \)|^2,
\end{align}
where the dimensionless power spectrum is defined in the usual way
\begin{align}
\langle \gamma_{\bf k}\gamma_{\bf k'} \rangle = (2\pi)^3\delta^{3}({\bf k}+{\bf k}')\frac{2\pi^2}{k^3}\Delta^{2}_{\gamma}.
\end{align}
Note that in deriving Eqn.\ (\ref{eqn:GWpower}) we have used the fact that the two contributions to the left handed modes in Eqn.\ (\ref{enhancement}) are not correlated with each other. The first term in this expression is the usual vacuum contribution, while the second term is the chiral enhancement due to the interaction with the gauge field fluctuations. In the lower panel of Fig.\  \ref{fig:analsol} we plot the ratio of the gravitational wave power in each polarization. 

\looseness-1

\subsection{Scalar perturbations} 
Let us also give a brief account of how scalar perturbations behave in this theory.
A suitable gauge choice is non-Abelian Coulomb gauge,
\be
\bar D_i\Psi_i = \partial_i\Psi_i -ig[A_i,\Psi_i] = 0,
\ee
which results in a Gauss law constraint that is algebraic in $k$-space.  This gauge condition can be used to eliminate $\chi_3$ and, combined with the solution to the Gauss law constraint, leads to an Lagrangian of the form
\be
\label{scalaract}
\mathcal{L}_S = \GG_{IJ}\Phi_I'\Phi_J' + \MM_{IJ}\Phi_I\Phi_J + \AA_{IJ}\Phi_I \Phi_J'
\ee
in terms of fields $\Phi \!=\! (\ada,\delta\!\phi , z)$, with $\ada \!=\! a\, \daxion$ the canonically normalized axion fluctuation and recall that $\delta\!\phi$ and $z$ are scalar components of the gauge field perturbations; here the coefficient functions $\GG$, $\MM$ and $\AA$ are rational functions of $x$ 
with coefficients that depend on $\mpsi$ and $\bflam$ (see \cite{Adshead:2013nka}). 
The action is that of a charged particle moving in three dimensions in a time-dependent harmonic well and magnetic field. 
The important terms in the action at large $\bflam$ and $x$ are a potential term $\bflam^2\mpsi^2\ada^2/x^2$ and ``magnetic'' coupling $(\bflam\mpsi/x)(z\!-\!\delta\!\phi)\ada'$.
In the asymptotic past $x\to\infty$, the mass matrix $\MM\to  \One$, and $\AA\to 0$, yielding free field equations for the modes.  
For $x \lesssim \bflam\mpsi$, the mode of interest is a ``magnetic drift'' mode where (as in the solution for the background) the large potential term balances against the large ``magnetic'' coupling, and the $\ada$ kinetic term is largely irrelevant.  A simplification of the dynamics ensues if we drop the $\ada$ kinetic term altogether, as well as the subleading $O(\bflam^0)$ terms in $\MM_{\axsub\axsub}$.  The axion dynamics is then determined in terms of the gauge scalars as
\be
\ada \!=\! 
\frac{(12 \mpsi^4 \!+\! 6 \mpsi^2 x^2 \!+\! x^4) \delta\!\phi + 2 x^4 z - 
 3 \mpsi^2 x (2 \mpsi^2 \!+\! x^2) \delta\!\phi' }{\bflam \mpsi^3 x^2}  .
\ee
Substituting this solution back into the action and making the further substitutions
\be
\delta\!\phi \to \frac{x(\mpsi  \phitil \!-\! x \ztil/\sqrt{3})}{3(2\mpsi^2+x^2)/\sqrt{2}}
,~
z \to \frac{ \mpsi x\phitil \!+\! (6\mpsi^2\!+\!x^2)\ztil/\sqrt{12}}{3(2\mpsi^2+x^2)/\sqrt{2}}
,
\ee
we find a reduced Lagrangian that depends only on $\mpsi$:
\bea
\label{Sred}
{\cal L}_{red} &=&  
 {\phitilb}'_k {\phitil}'_k+{\ztilb}'_k {\ztil}'_k
 -\frac{2}{\sqrt{3} {\mpsi}}( {\ztilb_k} {\phitil}'_k+ {\phitilb}_k {\ztil}'_k)
\nonumber \\
& & - \frac{(\mpsi^2-2)}{{\mpsi}^2} {\phitilb_k} {\phitil_k}
-\frac{2(\mpsi^2+1) }{\sqrt{3} {\mpsi} x} ( {\ztilb_k}{\phitil_k}+{\phitilb_k} {\ztil_k})
\nonumber \\
& &   \hskip 1cm
-\frac{\left(6 \mpsi^2+x^2+6\right) }{3 x^2} {\ztilb_k} {\ztil_k}  ~.
\eea
Note that the $\phitil$ dynamics is unstable when $\mpsi^2 \lesssim 2$.
In contrast to the tensor gauge modes, this scalar mode instability is present for a long time $\mpsi\lesssim x\lesssim\bflam\mpsi$, causing a breakdown of the linearized approximation.
Thus we discover what will prove to be an impediment to simultaneously satisfying all current observational bounds: small $\mpsi$ leads to unstable scalar modes, while large $\mpsi$ leads to unacceptably large tensor modes.  

\begin{figure}
\begin{center}
\includegraphics[width = 3in]{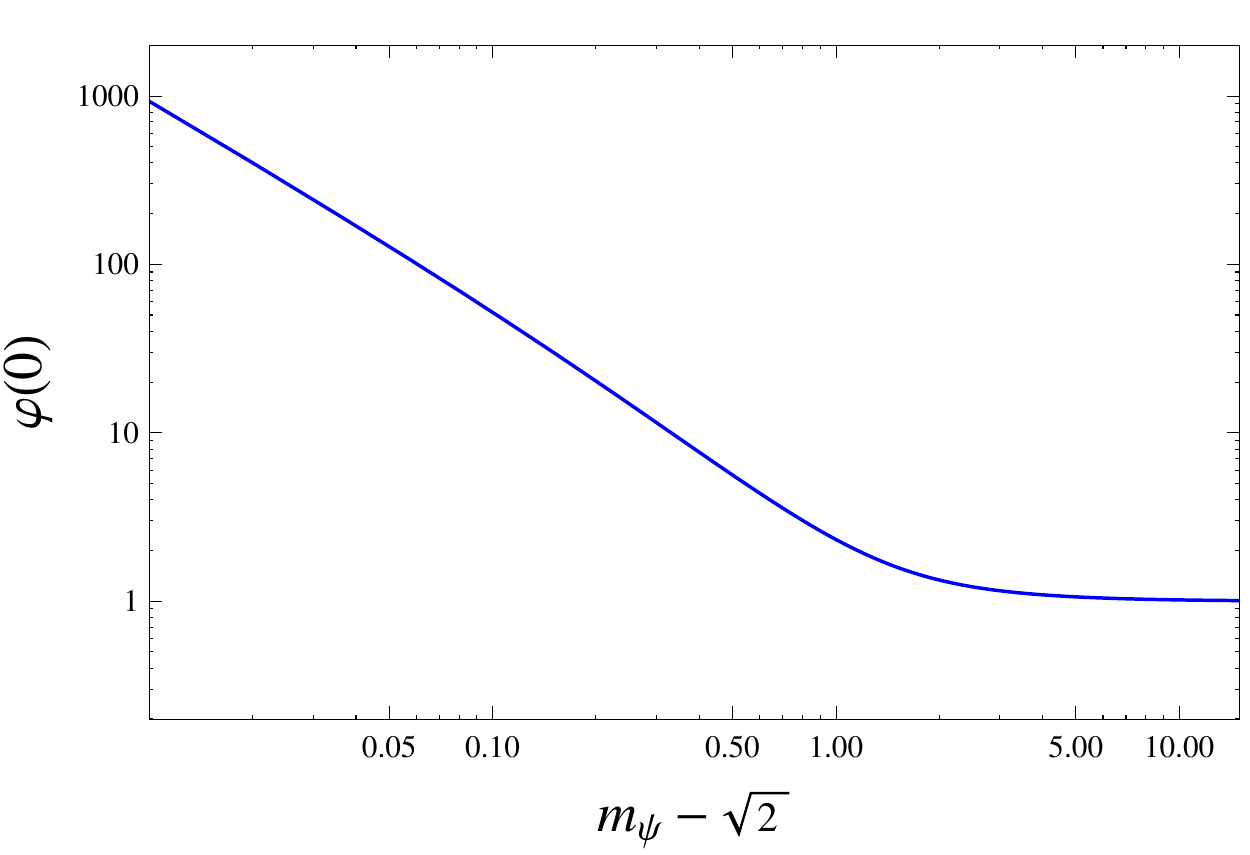}
\caption{ In this figure, we show the late-time value of $\phitil$ obtained from exact numerical integration of the reduced system as a function of $(\mpsi-\sqrt{2})$. }
\label{fig:phizero}
\end{center}
\end{figure}

The amplitude of the axion on super-horizon scales is determined (up to order one corrections) by the behavior of $\phitil$ near $x=0$ (with $k\neq 0$):
\be
\label{axionamp}
\ada \sim \frac{\sqrt{2}\, \phitil(0)}{\bflam\, x} + O(x)~.
\ee
The limiting value of $\phitil(x)$ as $x\to 0$ is of order one for large $\mpsi$, 
starts growing around $\mpsi \approx \sqrt{6}$,
and diverges as $\mpsi\to\sqrt{2}$ due to the softening potential for this mode; see Fig. \ref{fig:phizero}. 

We can now estimate the amplitude of scalar and tensor fluctuations. In spatially flat gauge, the curvature perturbation is ${\cal R} = H \delta u$, where $\delta u$ is the perturbation to the velocity potential, determined from
\begin{align}
\label{momentum}
T_{0i} 
\approx &~ (\bar{p}_{\axion}+\bar{p}_{\rm YM})N_i+a\dot\axion\partial_{i}\daxion\\ \nn
&  -\Bigl(H\psi \partial_{i}(2 z+4\delta\!\phi) - g a \psi^3\frac{\lambda}{f}\partial_{i}\daxion \Bigr)-2g^2\psi^4 N_i ~.
\end{align}
The term $\propto  g\psi^3 \lambda/f$ is much larger than the rest.
It comes from the gauge field momentum 
\be
T_{0i} = \tr(E\times B)_i \approx g^2 \tr( [A_j,\Psi_0][A_j,A_i] ) = ag^2\psi^3 \Psi_0^i
\ee
where 
$\Psi_0^i$ is determined by the Gauss law constraint
\be
\nabla\cdot E + \frac{\lambda}{f}\,\partial \axion\cdot B = 0 \ .
\ee
At late times, the leading piece of the first term comes from $g^2 [A_i,[A_i,\Psi_0]]$, and 
the leading piece of the second term is $(g\lambda/f) (\partial_i \daxion) [A_j,A_k] \epsilon^{ijk}$; 
thus one finds
\be
\Psi_0^i = \frac{\lambda}{gf} \partial_i \daxion \ ,
\ee
leading to the relevant term in~\pref{momentum}.
The scalar curvature perturbation that results is
\begin{align}
\curv \approx \frac{H}{\bar{\rho}+\bar{p}}g\psi^3 \frac{\lambda}{f}\frac{\hdaxion}{a} = \frac{g \psi^3}{2 H \epsilon_H}\frac{\lambda}{f}\frac{\hdaxion}{a} ~, 
\end{align}
where $\epsilon_{H} =  -\dot H/H^2$. 
The background has 
$\psi^3 \simeq -V_{,\theta} /(3 g \lambda H)$ and $\dot{H} \simeq  V_{,\theta}\,\dot \axion/(6fH)$, 
where $V_{,\theta}$ is the derivative of $V$ w.r.t. $\theta=\axion/f$,
thus
\begin{align}
\label{clock}
\curv \approx \frac{\daxion}{\dot \axion/H} ~.
\end{align}
Thus the axion is acting as the clock, as anticipated in \cite{Adshead:2012kp}, and the perturbation analysis is simply finding the magnitude of this perturbation.
Note that the magnitude of the perturbation, $\daxion$, was naively estimated in \cite{Adshead:2012kp} to be ${\cal O}(H)$ as in usual single-clock inflationary models.
As the calculation sketched before shows, however, that estimate was mistaken. As we can see from the discussion preceding Eqn. \ref{axionamp}, the multi-field scalar dynamics play a large role in determining the final amplitude for the axion perturbation; this is encoded in the function, $\phitil(0)$, and an amplitude decay factor,
$\bflam$. The upshot of this, relative to usual inflationary result for scalars on de Sitter, is that the magnitude of the $\daxion$ perturbation is reduced from $H$ by a factor of $\bflam$, up to the increase due to $\phitil(0)$. 
This decay in the amplitude of $\daxion$ compensates the reduction in the axion velocity by the same factor of $\bflam$ due to ``magnetic drift''. Hence, in contrast
with Ref. \cite{Adshead:2012kp},  we do not need to invoke very low values of $H$ to produce acceptable curvature perturbations, even when $\phitil(0)\sim1$.

\begin{figure}[t] 
   \centering
   \includegraphics[width=0.45 \textwidth]{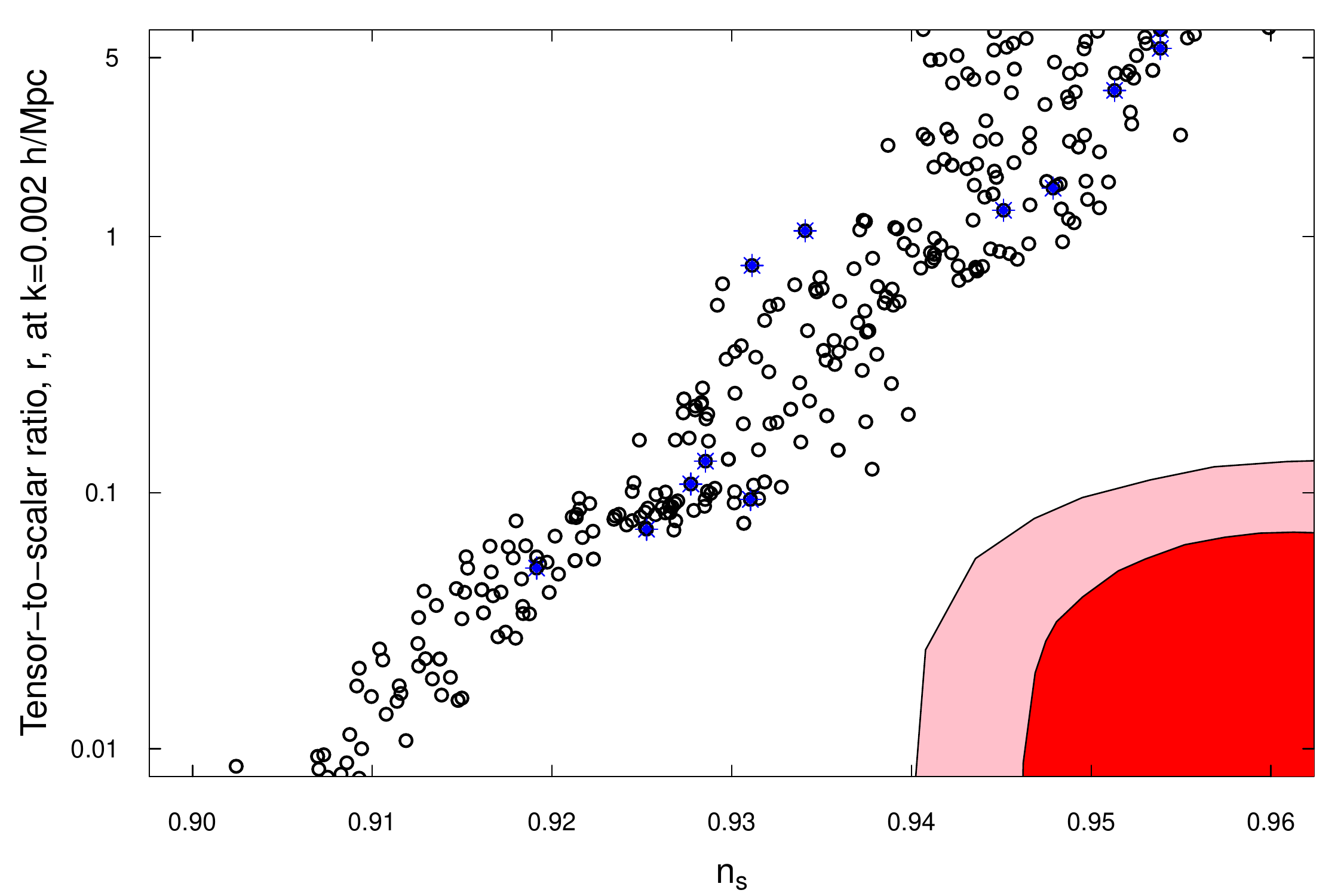} 
   \caption{Comparison of the tensor-to-scalar ratio, r, evaluated at $k=0.002$ h/Mpc and the spectral tilt, $n_s$ (evaluated at $k=0.05$ h/Mpc) for models
    drawn from a numerical exploration of the $g, f, \mu, \lambda$ parameter space with the corresponding parameter constraints from Planck. The value for $r$
    presented here includes the contributions from both gravitational wave helicities and is computed numerically using the gravitational
    wave mode functions. The open
   black circles represent parameter combinations whose scalar power spectrum amplitudes are outside of the Planck error bars; blue stars represent models
   with acceptable power spectrum amplitudes. The Planck one and two sigma contours are plotted in red and pink, respectively.
   Note that the y-axis is logarithmic, and that in this model it is possible to have $r>1$ due to the chirally enhanced gravitational wave spectrum.}
   \label{nsvr}
\end{figure}

Putting it all together, we find that the scalar curvature fluctuation is given by 
\be
\label{PsubR}
{\cal R} \simeq 
\frac{1}{\sqrt{2 k^3}}\cdot
\frac{H \mpsi  \phitil(0)}{\sqrt{2} (1+ \mpsi^2)} \Bigl(\frac{\lambda \mpsi V}{-V_{,\theta}}\Bigr)^{\!1/2}  ~;
\ee
(where in the reduced system, the scalar d.o.f.s are normalized to unit amplitude in the far past). We use
\bea
\label{slowrollparams}
\epsilon_H &=&- \frac{\dot{H}}{H^2}= -\frac{(1+\mpsi^2)V_{,\theta}}{\lambda\mpsi V},
\\
\nonumber
\eta_H &=& \epsilon_H+\frac1{2}\frac{ \dot\epsilon_H}{H \epsilon_H}
=  - \frac{(\mpsi^2+1)V_{,\theta\theta}}{\lambda\mpsi V_{,\theta}}  - \frac{(\mpsi^2-1)V_{,\theta}}{\lambda\mpsi V}  ~,
\eea
(recall $\theta \equiv \axion/f$) to write the curvature perturbation as
\be
\label{Reps}
{\cal R} \simeq
\frac{1}{\sqrt{2 k^3}}\cdot
\frac{H}{\sqrt{2\epsilon_H}}\cdot \frac{\mpsi \phitil(0)}{(1+\mpsi^2)^{1/2}} ~.
\ee
Relative to the corresponding expressions for single-field inflation, the slow-roll parameters~\pref{slowrollparams} directly exhibit the additional suppression of motion along the inflaton potential by the coupling $\lambda$ to the gauge sector; 
and in the perturbations, there is an additional enhancement from $\phitil(0)$ when $\mpsi$ is small enough.
Differentiating~\pref{Reps} with respect to $dN = -H dt$, one finds the spectral index
\be
n_s -1 \simeq 
-2\epsilon_H + \eta_H + 2 d\log\phitil(0)/dN ~.
\ee
The tensor-to-scalar ratio for the unenhanced graviton is
\be
r_- \simeq 
 \frac{8(1+\mpsi^2)\epsilon_H}{\mpsi^2\phitil(0)^2} ~,
\ee
while the other tensor-to-scalar ratio $r_+$ is enhanced by the effect discussed above, determined from Eqn.~\pref{eqn:GWpower}. At low values of $\mpsi$ the left-handed mode contributes equally, resulting in a factor of 2 in this expression. However, at large values of $\mpsi$, the gravitational wave power is dominated by the second term in Eqn.\ (\ref{eqn:GWpower}). To take into account the chiral enhancement, we solve the full perturbative dynamics numerically and perform a large exploration of parameter space.
\section{Discussion}
 We find that the model cannot simultaneously match all current observational bounds, as we summarize in Fig. \ref{nsvr}. This result can be schematically understood as follows. The difficulty for the model comes from the redness of its spectral tilt, $n_s$, which in this model gets contributions both from the slow roll parameters, $\epsilon_H$ and $\eta_H$ and from the gauge field dynamics through $d \log \phitil(0)/d N$. The slow roll parameters alone typically yield $n_{s}-1 \sim -0.03$ for this model. Since the gauge field contribution to the tilt is always negative, the background evolution -- encoded in the slow roll parameters -- leaves very little room for any extra tilt coming from $d \log \phitil(0)/dN$. Examination of Fig.\ \ref{fig:phizero} then suggests that, at low values of $\mpsi$, small variation of the gauge field mass parameter during the observable window translates to a large contribution to the tilt from this term. Thus, in order to have an acceptable tilt, it appears that one needs to arrange for a gauge field mass $\mpsi \gtrsim \sqrt{6}$ in order to sit on the plateau on the right side of  Fig.\ \ref{fig:phizero}. Turning now to the tensor to scalar ratio, it is immediately clear that values of
 $\mpsi$ this large will be a problem, because of the large chiral enhancement to the gravitational waves that such large $\mpsi$ values will lead to. In these models, the unenhanced tensor to scalar ratio is naturally large because the background inflationary mechanism mimics large field inflation, with  $\epsilon_H \sim 10^{-2}$, even though the field rolls over a very small distance in field space. This provides a counterexample to the most naive formulation of the Lyth bound. Unfortunately, this means that only small chiral enhancements of the gravitational waves on top of the ordinary vacuum amplitude can be tolerated by current observational constraints. For values of $\mpsi$ that yield acceptable scalar spectral indices, the tensor to scalar ratio is unacceptably large.
\looseness-1

%
{\bf Acknowledgements:} 
 We thank Wayne Hu for useful discussions and David Seery for early collaboration. This work was supported in part by DOE grant DE-FG02-90ER-40560 and by the Kavli Institute for Cosmological Physics at the University of Chicago through grants NSF PHY-1125897 and an endowment from the Kavli Foundation and its founder Fred Kavli. 
\looseness-1
{\sl Note Added:} 
While we were writing up our results, parallel work appeared~\cite{Dimastrogiovanni:2012ew} 
independently deriving the stability condition
$\mpsi^2>2$ 
using a WKB method.
\bibliography{ChiralGWLetter}
\end{document}